\documentclass{PoS}
\usepackage{graphicx}
\usepackage{array,booktabs,tabularx}
\usepackage{hhline}
\usepackage[numbers]{natbib}

\title{Long-Term X-ray Spectral Variability of Seyfert Galaxies with Swift}

\ShortTitle{Long-Term X-ray Spectral Variability of Seyfert Galaxies with Swift}

\author{\speaker{S. D. Connolly}\\
        University of Southampton, Highfield, Southampton, SO17 1BJ, UK\\
        E-mail: \email{sdc1g08@soton.ac.uk}}
        
\author{I. M. M\parbox[b][2.6mm][t]{2.0mm}{c}Hardy\\
        University of Southampton, Highfield, Southampton, SO17 1BJ, UK\\
        E-mail: \email{imh@astro.soton.ac.uk}}
 
\author{T. Dwelly \\
	Max-Planck-Institut f\"{u}r Extraterrestrische Physik, Giessenbachstrasse 1, 85748, Garching, DE\\
	E-mail: \email{dwelly@mpe.mpg.de}}

\abstract{
We present analysis of the long-term X-ray spectral variability of Seyfert galaxies as observed by {\it Swift}, 
which provides well-sampled observations over a much larger flux range and a much longer timescale
than any other X-ray observatory. We examine long-term variability of three AGN:
NGC 1365 (see Connolly et al. 2014), Mkn 335 and NGC 5548. At high fluxes, the 0.5-10 keV spectra soften with increasing flux,
as seen previously within the 2-10 keV band. However, at very low fluxes the sources also become very soft. We have fitted a 
number of models to the data and find that both intrinsic luminosity variability and variable absorption are required to explain the observations.
In some systems, e.g. NGC 1365, the best explanation is a two-component wind model in which one component represents direct
emission absorbed by a disc wind wind, with the absorbing column inversely proportional to the intrinsic luminosity, and the second
component represents unabsorbed emission reflected from the wind. In other AGN the situation is more complex.}

\FullConference{Swift: 10 Years of Discovery,\\
		2-5 December 2014\\
		La Sapienza University, Rome, Italy }

\begin{document}

\section{Introduction}
\label{intro}

\vspace{-2mm}
X-ray spectral observations have shown that many Seyfert galaxies possess absorbers whose absorbing columns vary by large amounts  \citep{Risaliti2002}. 
The detection of variability of the absorbing material on a timescales of hours has indicated that it must be close
to the nucleus, at a distance similar to that of the Broad Emission Line Region \citep[e.g.][]{Puccetti2007}; some absorption is claimed to be by clouds within the Broad
Line Region, due to complete occultations which have been observed on timescales of days \citep{Risaliti2007a}. The existence of absorbing outflows in Seyferts,
linked to disc winds, is also well established observationally \cite{Tombesi2013}.
Longer-term trends in the behaviour of these absorbers have, however, not been investigated
in as much detail.

Here, we look at the long-term variability of three Seyfert galaxies - NGC 1365, NGC 5548 and Mkn 335 - which are all known to possess absorption variability. 
The three systems cover a range of masses, Eddington ratios and viewing angles (see Table \ref{paramsTable}).

\begin{table}[ht!]
    \label{paramsTable}
    \centering
    \small
	    \begin{tabular}{ l c c c } \hline			   
	       & Black Hole Mass $\mathrm{^{[4,5,6]}}$ ($\mathrm{M_{\odot}}$) & Eddington Ratio ($\mathrm{L_{Bol}/L_{Edd}}$)$\mathrm{^{[7]}}$ & Viewing angle $\mathrm{^{[8]}}$ ($^{\circ}$) \\ \hline
    NGC 1365   & $\mathrm{2.9\times10^7}$		  & $\mathrm{0.02}$			 & $\mathrm{36-60}$	\\ \hline
    NGC 5548   & $\mathrm{6.5\times10^7}$		  & $\mathrm{0.04}$			 & $\mathrm{22-40}$	\\ \hline
    Mkn 335    & $\mathrm{2.5\times10^7}$		  & $\mathrm{0.38}$			 & $\mathrm{38-49}$	\\ \hline
    \end{tabular}
    
    \caption{Physical parameters of each of the three Seyfert galaxy nuclei.}
    
	\vspace{-3mm}
\end{table}

\vspace{-6mm}

\section{Observations \& Data Reduction}
\label{obs}
\vspace{-2mm}
The entire set of {\it Swift} XRT data available on the HEASARC archive$^1$\footnotetext[1]{http://heasarc.gsfc.nasa.gov/cgi-bin/W3Browse/swift.pl} 
for each source was used for this study. In each case, 
several hundred spectra from individual {\it Swift} `visits', or exposures, were used, each with a total of several hundred kiloseconds of exposure time. 
A small fraction of the spectra were rejected due to very low signal to noise or artifacts near the source.

The data were reduced with an automatic pipeline which uses the XRTPIPELINE software, previously employed in e.g. \cite{Connolly2014}. 
The XSELECT tool was used to extract spectra and lightcurves, using flux-dependent source and background extraction regions chosen such that background contamination was reduced at 
faint fluxes and the effects of pile-up were mitigated at high fluxes. The X-ray background was subtracted using a background annulus region.
The observed XRT count rates were then corrected for bad pixels, vignetting effects, and the finite extraction aperture.

\begin{figure}
	\centering
	\includegraphics[width = 0.9\columnwidth,trim = 0 440 0 0, clip=true]{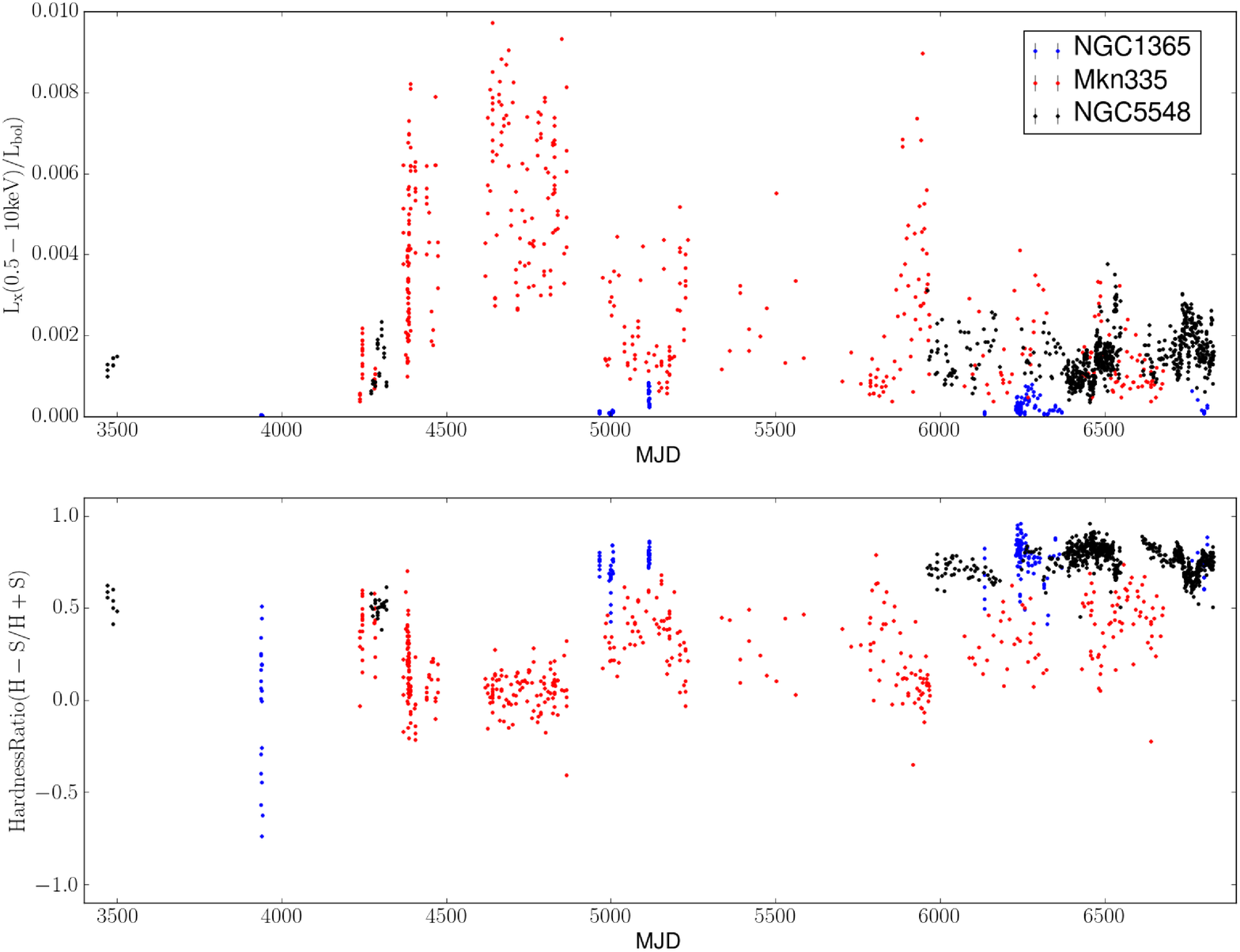}
		
	\caption{The ratio of 0.5-10 keV X-ray luminosity to Eddington luminosity of each source, as a proxy for the Eddington ratio.}
	\label{lc}
\end{figure}

Fig. \ref{lc} shows a lightcurve of all of the {\it Swift} data over the nine-year period of observation.

\vspace{-4mm}

\section{Data Analysis}
\vspace{-2mm}
\subsection{Spectral Hardness}
\vspace{-2mm}
\begin{figure}[ht!]
	\centering
	\includegraphics[width = \columnwidth ,trim = 0 0 0 0, clip=true]{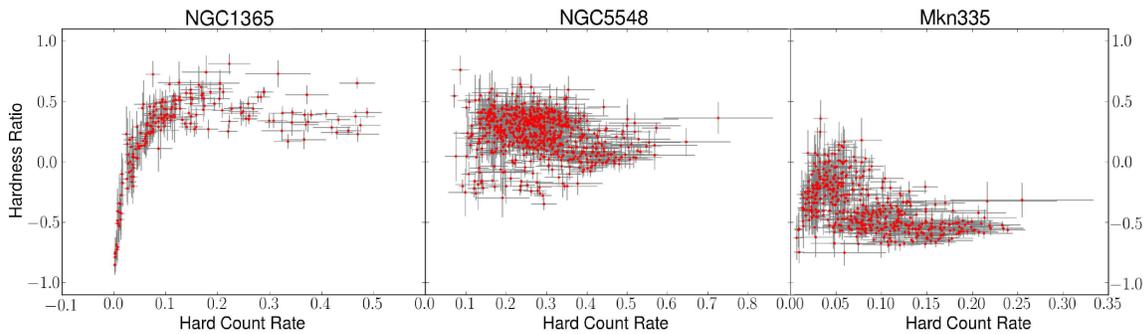}
	
	\caption{The hardness ratio (H-S/H+S) against hard count rate for each source. }
	\label{hardness}	
\end{figure}

Fig. \ref{hardness} shows the hardness ratio (H+S/H-S) against the hard count rate of each source. 
Here, we look at spectral variations within the $0.5-10.0$ keV band, a broader band than the $2.0-10.0$ keV energy band on which most previous studies concentrate. 
As the low-energy end of the spectrum is more sensitive to absorption, this energy range enables a more complete look at spectral changes due to absorption. 
In each case, hard and soft emission are defined as $2.0 - 10.0$ keV (H) and $0.5-2.0$ keV (S), respectively. 
 
The hardness ratios all show similarities - all three display softening with increasing flux at higher flux levels. 
NGC 1365 also becomes extremely soft at the lowest flux levels. The Mkn 335 data hint at a similar trend, but are more similar 
to those of NGC 5548. 

NGC 1365 shows a very small amount of scatter in hardness, implying that the spectral shape is fairly constant at a given flux level. 
The spectra of NGC 1365 could therefore be binned by flux, giving a
high signal-to-noise ratio spectrum at each flux level. As the Mkn 335 and NGC 5548 data show much greater variation in hardness at a given flux,
they were instead binned in time intervals.

\subsection{Spectral Modelling}

\begin{figure}[ht!]
	\centering
	\includegraphics[width = 0.5\columnwidth, angle = 0,trim = 0 0 0 0,clip = true]{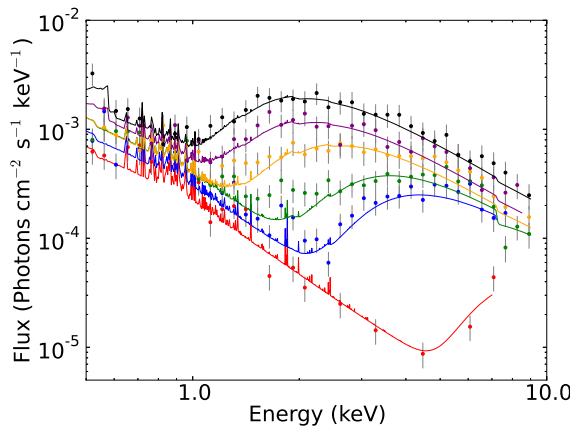}
	\caption{Sample of the flux-binned spectra of NGC 1365, 
	simultaneously fitted with the best-fitting model, consisting of two power laws, 
	one of which is absorbed by a partially ionised absorbing column.}
	\label{spectra}
\end{figure}

The binned spectra of each source were fitted with two possible absorption models - a partial covering model and a two-component model consisting of two
power laws, one of which is absorbed by an ionised absorbing column (e.g. Fig \ref{spectra}).  
Table \ref{fits} shows the reduced $\chi^2$ values for variants of each model.
Both models have previously been used to describe AGN spectra, e.g. NGC 4945 \cite{Done1996} and NGC 1365 \cite{Risaliti2009}.
In all fits, the underlying power law spectral index is tied, such that it remains constant between spectra. In reality, $\Gamma$  
is likely to vary slightly, and allowing it to vary does give better fits. However, due to a degeneracy with varying absorption, a large, 
unphysical range of $\Gamma$ is produced.

\begin{table}[ht!]

\centering
\footnotesize
\begin{tabular}{| c || c | c | c || c | c | c |} \hline	

	 & \multicolumn{3}{c}{Absorbed \& unabsorbed powerlaws} 	& \multicolumn{3}{c}{Partial Covering} 			\\ \hhline{~------}
	 & All free &	$\mathrm{N_H}$ tied  	& Ionisation Tied	& All free 	& $\mathrm{N_H}$ tied	& Covering fraction tied\\ \hline
NGC 1365 & 1.39	    &   2.28			& 1.42			& 1.45		& 2.57			& 1.45			\\ \hline
NGC 5548 & 1.03	    &   1.13			& 1.04			& 1.06		& 1.14			& 1.09			\\ \hline
Mkn 335  & 1.12	    &   1.13			& 1.14			& 1.24		& 1.23			& 1.35			\\ \hline

\end{tabular}

\caption{Total reduced $\chi^2$ values of variations of two absorption models, fitted to the average spectra of each source.}
\label{fits}
\end{table}

The best-fitting model in all cases was the two-component model. The best-fit parameters showed that the variability of NGC 1365 requires a decrease in absorbing 
column with increasing flux, regardless of whether or not the ionisation state was allowed to vary - changes in the ionisation state of the absorber alone were found to be
unable to account for the observed spectral variability. When both the absorbing column and ionisation state were allowed to vary freely, only small changes in the ionisation 
were required by the best-fitting model
and the inverse relationship between the absorbing column and flux was still present. In Mkn 335 and NGC 5548, however, changes in ionisation state and absorbing column
alone described the spectral variability equally well, meaning neither source of variability could be ruled out.


\begin{figure}[ht!]
	\centering
	\includegraphics[height = 0.35\columnwidth,trim = 0 0 0 0, clip=true]{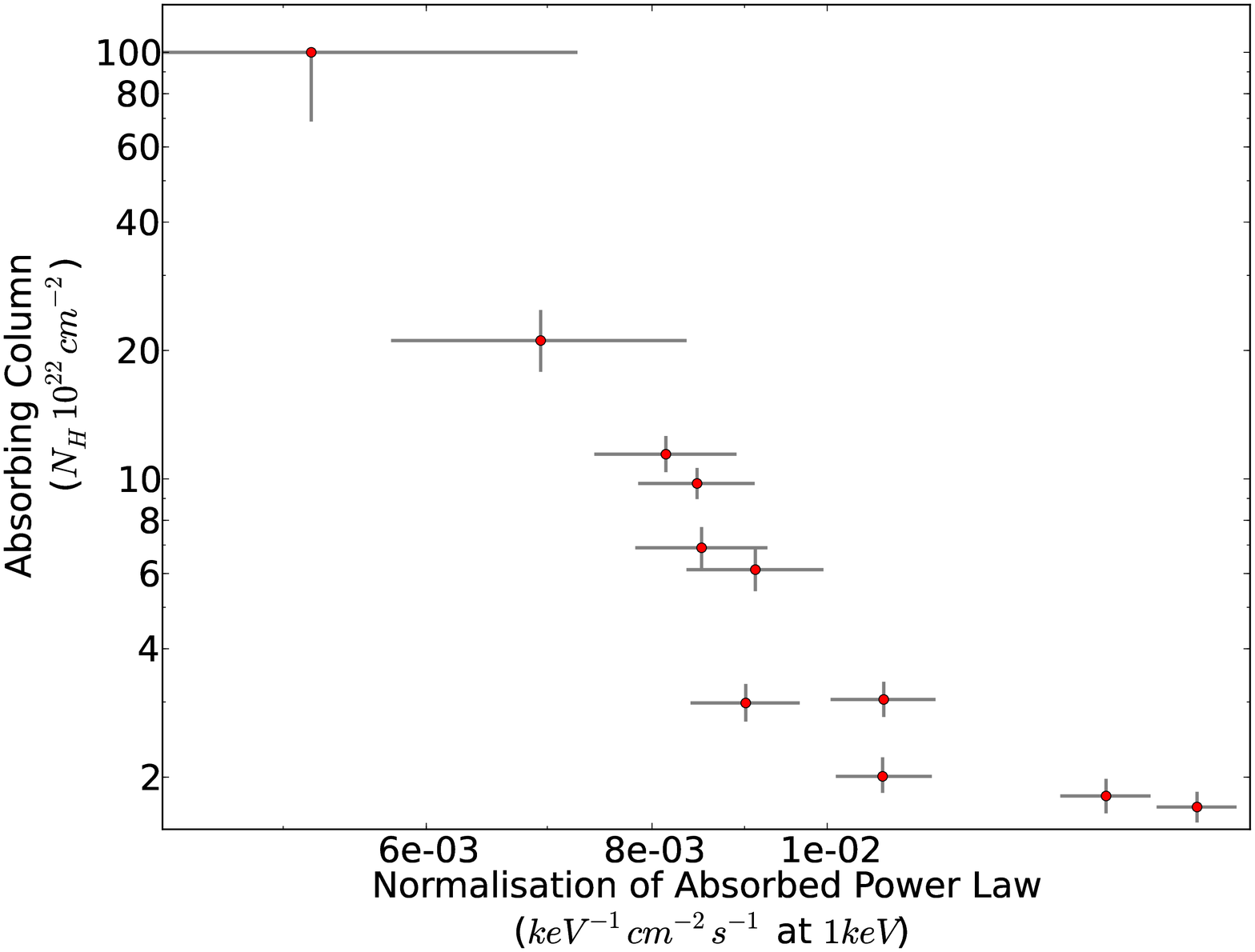}\hspace{1pt}
	\includegraphics[height = 0.37\columnwidth,trim = 0 0 10 0, clip=true]{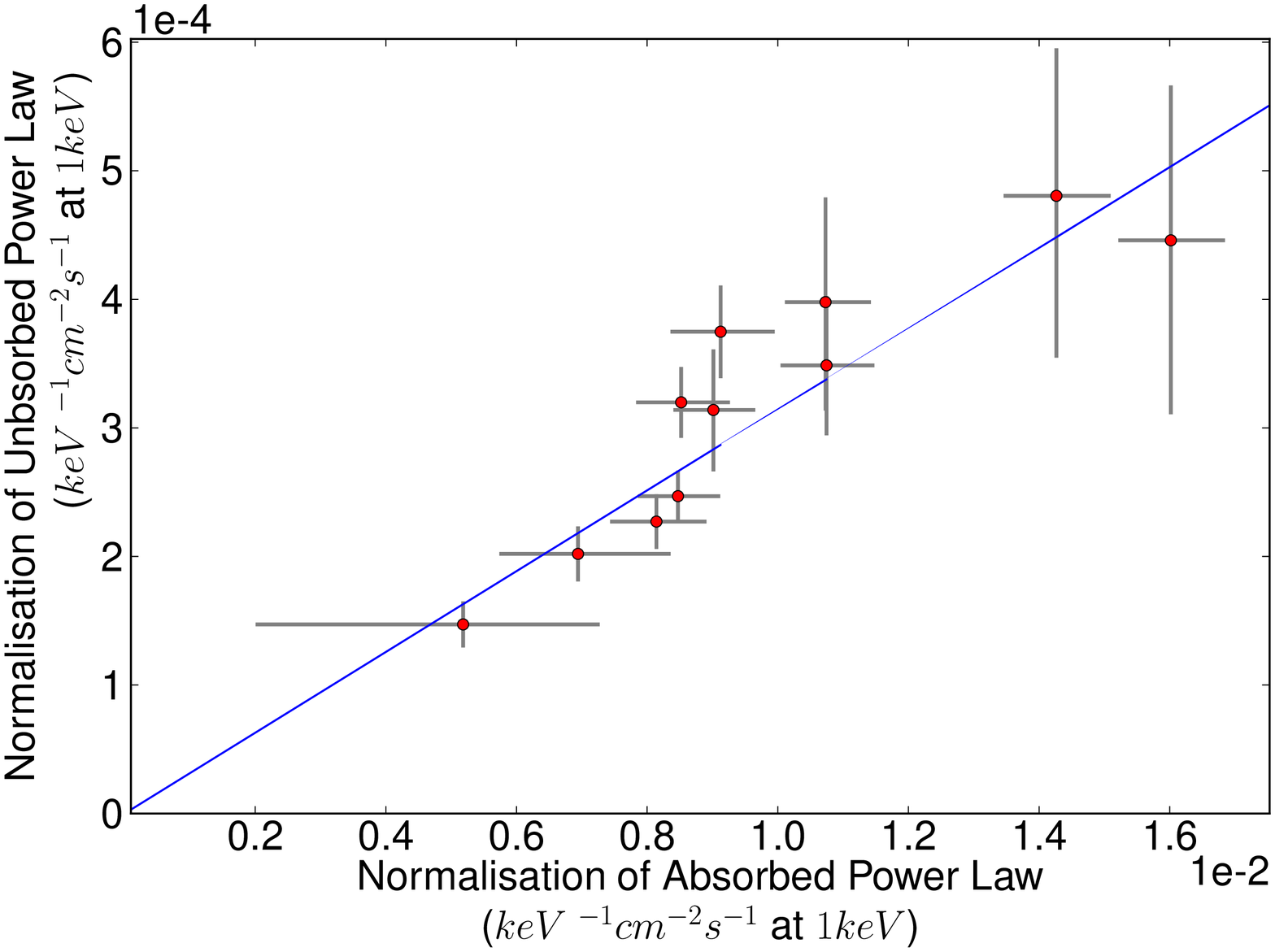}\hspace{1pt}
	\caption{Plots of the parameters of the two-component model of NGC 1365.
	(Left) The normalisation parameter of the absorbed power law against the column of the absorbing material. 
	(Right) The normalisation parameter of the absorbed power law against that of the unabsorbed power law, fitted with a linear model. }
	
	\label{nhplots}
\end{figure}  
				



\section{Discussion}

\vspace{-2mm}
\subsection{A Possible Relationship Between Source Flux and Column Density}

Fig. \ref{nhplots} (left) shows a clear decrease in the absorbing column as the normalisation of the absorbed power law increases in the two-component model of NGC 1365. 	
As this component dominates the unabsorbed luminosity, the absorbing column is therefore inversely proportional to the source luminosity. 
Whilst a reduction in absorption with increasing flux might initially be assumed to be
due to increased ionisation, spectral models involving varying ionisation alone do not fit the data. 
Fits to spectra of NGC 4151 at different flux levels show a similar reduction in the absorbing column with increasing flux \citep{Lubinski2010}, implying
that this relationship is not unique to NGC 1365. 

For the same model, Fig. \ref{nhplots} (right) shows the normalisation of the unabsorbed power law to be well correlated with that of the absorbed power law (r = 0.91), implying a common origin. 
The underlying luminosity of the source is also shown to be changing; the observed changes in flux are therefore 
not solely due to changes in the absorbing medium. 

A similar inverse relationship between the absorbing column and source luminosity is seen in the two-component model
of Mkn 335 in which the ionisation state is not allowed to vary. However, the data are equally well fit when the ionisation state alone is allowed to vary,
meaning the inverse relationship cannot be confirmed. As the spectra of NGC 5548 are best fit by models in which only the absorber's ionisation state varies,
the mechanism of absorption variability is likely to be different to that of NGC 1365. Both NGC 5548 and Mkn 335 also show underlying continuum variability in addition
to absorption variation.

Although unlikely to explain the absorption variability of all AGN, the spectral variability displayed by at least one 
AGN can be entirely explained by systematic variation of the absorber, with the absorption varying inversely with luminosity.

\subsection{AGN Wind Model}

\begin{figure}[ht!]
	\centering
	\includegraphics[width = 0.6\columnwidth,trim = 0 0 0 0, clip=true]{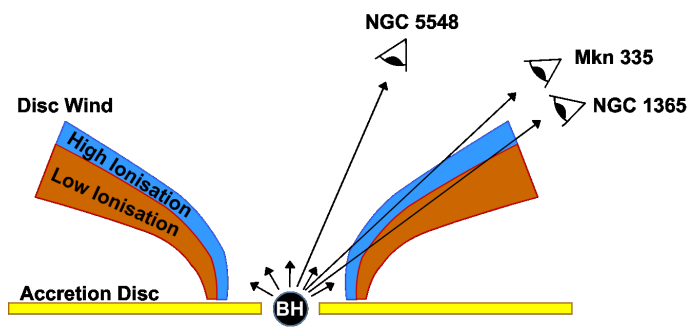}	
	\caption{Possible orientations of each AGN with respect to a disc wind. As in the \citet{Elvis2000} wind model.}
	\label{windangles}	
\end{figure}
\vspace{-2mm}
Absorption in AGN X-ray spectra is often attributed to disc winds (see e.g. \cite{Kaastra2000}, \cite{Tombesi2013}). 
Both models and single-epoch data samples suggest that a rise in accretion rate, 
and therefore an increase in X-ray luminosity, naturally leads both to an increase in the radii from which the wind arises,
and to an increase in the opening angle of the wind \citep{Nicastro2000, Elvis2000,Tombesi2013}.   
Thus, if the observer views the X-ray source at an inclination such that they are looking through the inner edge of the wind, an increase in accretion rate would move the wind out of the
line of sight of the observer, lowering the measured absorbing column. This  will thus naturally lead to the inverse relationship between
absorbing column and luminosity which we see clearly in NGC 1365. If Mkn 335 is observed through the very inner part of the wind, such that ionisation effects are greater,
whilst our view of NGC 5548 is unobscured (see Fig. \ref{windangles}), this model has the potential to explain the variability of all three AGN.

\vspace{-2mm}
\section{Conclusions}
\vspace{-2mm}
The {\it Swift} spectra of NGC 1365, Mkn 335 and NGC 5548 show large spectral variation, 
a significant cause of which is found to be changes in the absorbing material.
In NGC 1365, the absorbing column decreases with increasing X-ray flux,
which could be explained by a varying disc wind. The Mkn 335 data hints at
similar behaviour at lower fluxes, whilst the variation of NGC 5548 is best explained by
changes in the ionisation state of the absorber.
\vspace{-2mm}
\section*{Acknowledgments}
\vspace{-2mm}
SDC thanks the STFC for support under a studentship and IMcH thanks the STFC for support via grant ST/G003084/1. 
\vspace{-2mm}

\end{document}